\documentclass[reprint,superscriptaddress,pre,amsmath,amssymb,aps]{revtex4-2}

\usepackage[utf8]{inputenc}  
\usepackage[T1]{fontenc}
\usepackage{xcolor}
\usepackage{graphicx}
\usepackage{url}
\usepackage{orcidlink}

\begin{document}

\preprint{APS/123-QED}

\title{On the Effects of Decentralized Moderation on Network Robustness and Information Diffusion in Mastodon}
\author{Beatriz Arregui-Garc\'ia\,\orcidlink{0000-0002-5591-8367}}
\affiliation{Fondazione Bruno Kessler, Via Sommarive 18, 38123 - Povo, Italy}
\author{Lucio La Cava\,\orcidlink{0000-0003-3324-0580}}
\affiliation{Department of Informatics, Modeling, Electronics and System Engineering
University of Calabria, Via P. Bucci 44Z, 87036 - Rende, Italy}
\author{Anees Baqir\,\orcidlink{0000-0002-5771-1933}}
\affiliation{Department of Computing, Northeastern University London, London, UK}
\author{Andrea Tagarelli\,\orcidlink{0000-0002-8142-503X}}
\affiliation{Department of Informatics, Modeling, Electronics and System Engineering
University of Calabria, Via P. Bucci 44Z, 87036 - Rende, Italy}
\author{Riccardo Gallotti\,\orcidlink{0000-0002-8088-1973}}
\affiliation{Fondazione Bruno Kessler, Via Sommarive 18, 38123 - Povo, Italy}
\author{Sandro Meloni\,\orcidlink{0000-0001-6202-3302}}
\email[]{sandro@ifisc.uib-csic.es}

\affiliation{Institute for Cross-Disciplinary Physics and Complex Systems (IFISC), CSIC-UIB, 07122 - Palma de Mallorca, Spain}
\affiliation{``Enrico Fermi'' Research Center (CREF), Via Panisperna 89A, 00184 - Rome, Italy}

\date{\today}

\begin{abstract}
Decentralized online social networks such as Mastodon distribute moderation power across thousands of independently governed servers, raising fundamental questions about how local block decisions shape global structure and information flow. In this paper, we analyze Mastodon at the instance level by constructing a signed, directed, temporal network in which positive edges aggregate inter-instance follow relationships and negative edges encode daily block actions. Using one year of data, we show that despite continuous moderation activity and changing roles among instances, the network exhibits strong structural stability: signed dyadic motifs and degree distributions display highly persistent dynamics, and aggregated transition matrices satisfy Markovian equilibrium conditions over intermediate time scales. Building on the marked asymmetry between instances that predominantly issue bans and those that are mostly banned, we then study information diffusion on the positive network via a hybrid contagion model that combines simple contagion within groups and complex contagion across groups. We find that information originating in the minority of moderating instances spreads more efficiently, both internally and toward the majority, while the opposite direction is fragile and sensitive to contagion parameters. Echo-chamber effects emerge even in a globally balanced signed network and become stronger under stricter contagion conditions. Together, these results show that decentralized moderation in Mastodon generates a stable macroscopic configuration that both structures and constrains information exchange, effectively isolating norm-violating domains without centralized control.
\end{abstract}

\maketitle

\section{Introduction}
Over the past two decades, online social networks (OSNs) have become essential infrastructures for communication, information exchange, and social organization \cite{LAZER2009CSS,BAKSHY2012CSS,MENG2025CSS,LORENZ2019CSS}. The scale at which millions of users—and increasingly, automated agents—participate in these environments generates complex, coordinated behavior \cite{DESIDERIO2025RISKS} that can pose significant risks to public health or electoral integrity \cite{GALLOTTI2020INFODEMICS,BOVET2019RISKS,GRINBERG2016FAKENEWS}. A wide range of harmful content, including misinformation, harassment, hate speech, and coordinated manipulation, can spread rapidly across densely connected systems, often outpacing mechanisms designed to detect or counteract it \cite{LAZER2018OSN,VOSOUGHI2018FAKENEWS, JUUL2021DIFFUSION}. Studies across multiple platforms have shown that online communities can evolve into echo chambers, where selective exposure, homophily, and social reinforcement jointly amplify extreme or misleading content \cite{DELVICARIO2016ECHO,CINELLI2021ECHO,CONOVER2021ECHO}.\\
Mainstream platforms have responded with various content-moderation strategies, ranging from algorithmic filtering and fact-checking to the permanent removal of offending accounts—a practice known as deplatforming \cite{JHAVER2021DEPLATFORMING,INNES2023DEPLATFORMING,ROGERS2020DEPLATFORMING}. While deplatforming can effectively reduce the immediate reach of harmful content \cite{THOMAS2023DEPLATFORMING}, it can also generate unintended consequences: banned users often migrate to alternative platforms, where they may form ideologically homogeneous communities and continue their activities with diminished oversight \cite{ALI2021DEPLATFORMING, Horta2023Parler}. 
These dynamics highlight a fundamental dilemma of centralized moderation: it can limit harmful exposure within a platform, but may simultaneously fragment public discourse and accelerate the proliferation of less-regulated spaces \cite{MEKACHER2023DEPLATFORMING}.\\\\

\noindent
As concerns about centralized governance grow, decentralized online social networks (DOSNs) have attracted increasing attention as an alternative paradigm. The structural properties of these networks have been studied with work focusing on the growth dynamics and backbone of the instance network \cite{LaCava2021FediverseGrowth}, and on how user communities are distributed across instances \cite{Brauweiler2025, Sabo2024MastodonAdoption}. In platforms such as Mastodon, content hosting, user data, and moderation responsibilities are distributed across independently operated servers—known as instances—interconnected through open protocols such as ActivityPub \cite{Rozenshtein2023Fediverse}. Rather than a single corporation enforcing uniform rules, each instance defines its own moderation policies, community norms, and governance practices. This federated design promises greater user autonomy, transparency, and resilience by eliminating a single point of control and enabling communities to self-organize according to their values \cite{Anaobi2023,Surve2024,BONO2024}. A growing body of research has examined moderation in the Fediverse, predominantly through qualitative methods. Interview-based studies have explored the challenges faced by instance administrators and moderators, including the labor involved in moderation \cite{SpencerSmith2025LabourPains} and the use of community-level blocklists \cite{Melder2025, Zhang2025CommunityBlocklists}. Other qualitative work has examined the political dimensions of federated governance \cite{Gehl10122023}, the formal rules adopted by Mastodon instances \cite{Nicholson2023MastodonRules}, and how governance structures and moderation priorities evolve as communities grow \cite{Muralidharan2026FederatingGovernance}. Quantitative research remains comparatively limited. Existing studies include an analysis of defederation events, showing that blocking reduces activity on targeted servers without significantly affecting post toxicity \cite{Colglazier2024GroupSanctions}, using LLMs for automating community rule compliance \cite{LaCava2025osnem}, and a federated graph-learning framework for detecting toxic conversations across instances while preserving data locality \cite{Leonidou2026DeToxFed}.\\

Despite growing attention to DOSNs, the mechanisms by which decentralized moderation shapes collective structure and information flow remain poorly understood. It remains unclear whether federated systems—subject to continuous, heterogeneous moderation activity—can settle into stable macroscopic configurations, or whether they exhibit persistent structural flux. While recent work has identified equilibrium-like properties in offline social ties \cite{MIGUEL2025EQUILIBRIUM}, comparable analyses in digital environments are scarce. Research on online networks has largely focused on growth mechanisms such as preferential attachment \cite{LESKOVEC2008NETWORKEVOL}  or bursty edge dynamics \cite{GAITO2012BURSTYGROWTH}, rather than on the possibility of stationary regimes. This question is especially pertinent in Mastodon, where moderation does not merely regulate behavior but actively rewires the network: instances can block others, severing communication channels and reshaping the system's topology. Moreover, structural features of signed networks—such as patterns of trust and distrust—have been shown to generate echo chambers and polarization even in the absence of algorithmic filtering \cite{VENDEVILLE2025BALANCEINDEX}, raising the further question of whether moderation boundaries shape not only network structure but also the efficiency and directionality of information transmission. We address both questions: whether decentralized moderation produces a stable macroscopic configuration, and whether that stable structure leaves a fingerprint on what information reaches whom.\\

\noindent
In this paper, we investigate how decentralized moderation shapes the structural stability and information flows of Mastodon's inter-instance network over one year of observation. We combine two large-scale datasets: a snapshot of inter-instance follow relationships and a daily record of instance-level domain blocks spanning August 2023 to July 2024. Representing the system as a signed, directed, temporal network—where positive edges encode social connectivity and negative edges encode moderation decisions—we show that, despite continuous moderation activity and evolving participant identities, the network's edge configurations and degree distributions display strong temporal persistence, satisfying equilibrium conditions at mesoscopic temporal scales. Building on the structural asymmetry between instances that predominantly issue bans and those that are primarily targeted, we then simulate information diffusion using a hybrid contagion model that combines simple contagion within groups (moderators and banned instances) and complex contagion across group boundaries. We find that information originating in the minority of moderating instances spreads more efficiently—both internally and toward the majority—while the reverse direction is fragile and sensitive to contagion parameters. Echo-chamber effects emerge even in a globally balanced signed network and intensify under stricter contagion conditions. Together, these results show that decentralized moderation in Mastodon generates a stable macroscopic configuration that both structures and constrains information exchange, effectively isolating norm-violating domains without centralized control.

\section{Network Description}

\begin{figure}
    \centering
    \includegraphics[width=0.5\textwidth]{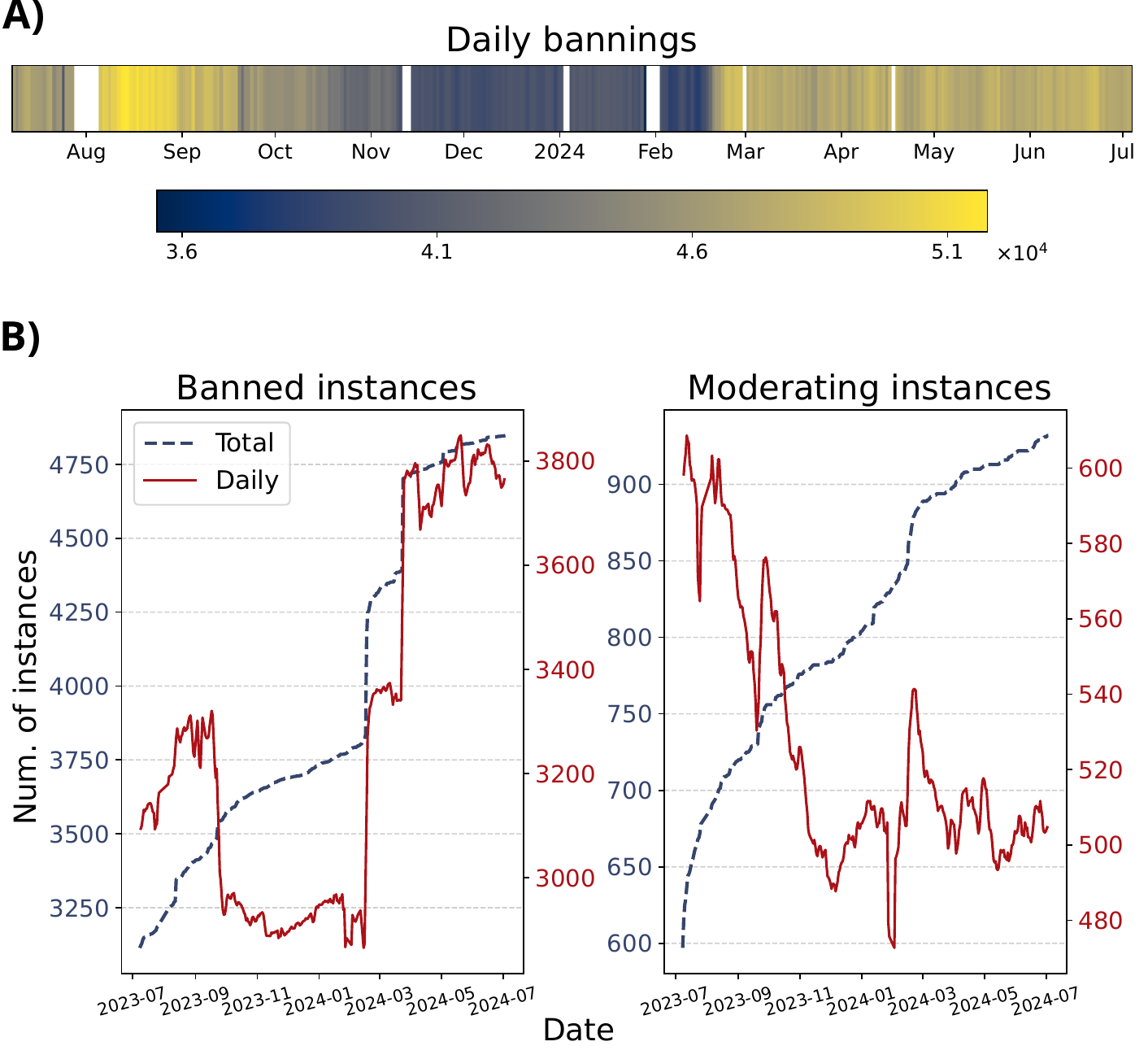}
    \caption{\textbf{Moderation activity in the Mastodon social network:} 
    Panel A shows the total number of bans registered across instances on the platform; white regions indicate days with missing data. 
    Panel B is divided into two subpanels: the left panel shows instances receiving bans, while the right panel shows instances issuing bans (moderators). 
    In both subpanels, the daily number of distinct instances being banned or acting as moderators is shown in red, while the cumulative number of these instances is shown as a dashed blue line.}
    \label{fig:mastodon_infobans}
\end{figure}

We represent the Mastodon ecosystem at the instance level as a signed, directed network capturing both social connectivity and moderation activity. The network is constructed by combining two complementary data sources: (i) a snapshot of inter-instance follow relationships, and (ii) a daily record of instance-level domain blocks spanning August 2023 to July 2024. Since these datasets originate from independent collection pipelines and cover partially overlapping sets of instances, we restrict our analysis to the induced subgraph of 5413 instances for which both types of information are simultaneously available. We refer to this network as the \emph{governance-social interface}, as it represents precisely the portion of the Mastodon federation where social connectivity and governance practices co-occur and interact. More details on data collection are provided in Supplementary Material.\\

\noindent
Let \(V^*\) denote the set of instances in the governance-social interface. Each node \(i \in V^*\) corresponds to a single Mastodon instance. Edges between instances carry a sign reflecting two qualitatively distinct forms of inter-instance interaction: domain-level moderation actions (negative) and follower relationships (positive), described below.\\

\paragraph{Moderation network.}
Negative ties originate from domain block actions recorded in daily blocklists over the one-year observation period. On Mastodon, domain blocks are enacted by instance administrators and apply to all accounts hosted on the targeted instance, making them a collective, institutional form of inter-instance regulation. We represent the moderation layer as a time-dependent directed graph
\[
G^{-}(t) = (V^*, E^{-}(t)),
\]
where a directed edge \((i,j) \in E^{-}(t)\) indicates that instance \(i\) issued at least one domain block targeting instance \(j\) on day \(t\).\\

\noindent
Moderation activity is substantial and sustained throughout the observation period as illustrated in Fig. ~\ref{fig:mastodon_infobans}. We record an average of approximately \(4.5 \times 10^4\) block actions per day, peaking at around \(5.2 \times 10^4\) in August 2023, dropping to a minimum of approximately \(3.6 \times 10^4\) during winter 2023--2024, and stabilizing near \(4.7 \times 10^4\) for the remainder of the period. A marked asymmetry characterizes the actors involved: on any given day, the number of distinct instances being targeted substantially exceeds the number of instances issuing blocks. Over the full observation period, we identify approximately 932 unique moderating instances and 4,848 unique moderated instances. Both populations display continuous turnover, with new actors appearing over time in both roles, indicating that moderation dynamics are sustained by an evolving set of participants rather than a fixed core of recurrent agents as we can see in Supplementary Fig.~1 and Supplementary Fig.~2.\\

\paragraph{Follow network.}

Positive ties are derived from user-level follower relationships aggregated to the instance level. Given two instances \(i\) and \(j\), we define a directed positive edge \((i,j) \in E^{+}\) if at least one user hosted on instance \(i\) follows at least one user hosted on instance \(j\). The follow network is thus represented as a weighted directed graph
\[
G^{+} = (V^*, E^{+}, W^{+}),
\]
where the weight \(w_{ij}^{+}\) corresponds to the total number of cross-instance follower relationships from \(i\) to \(j\).\\


\paragraph{Signed temporal network.}
Combining both layers, we represent the full system as a signed weighted temporal network
\[
G(t) = (V^*, W(t)),
\]
where the signed adjacency matrix \(W(t) = \{w_{ij}(t)\}\) encodes both interaction types. For each ordered pair \((i,j)\), \(w_{ij}(t) > 0\) if instance \(i\) maintains follower connections toward \(j\) (with magnitude equal to the number of cross-instance follower relations), \(w_{ij}(t) = -1\) if instance \(i\) issues a domain block targeting \(j\) on day \(t\), and \(w_{ij}(t) = 0\) otherwise. Although the positive layer is static, the network as a whole is time-dependent due to the evolving structure of moderation interactions.\\

\noindent
During the initial months of observation, approximately 16\% of active inter-instance links are negative. This proportion decreases over time and stabilizes around 10\% during the final four months, indicating a relative decline in the density of moderation ties compared to the persistent affiliative backbone.

\subsection{Network dynamics}
To evaluate the temporal stability of the signed network, we analyze the evolution of dyadic edge configurations over time. Each ordered pair of instances \((i,j)\) can occupy one of nine possible states depending on the sign of the edge in each direction:
{
\[
\{(-1,-1), (-1,0), (-1,1), \dots, (1,-1), (1,0), (1,1)\}
\]
}

\noindent
encoding all combinations of moderation (\(-1\)), absence of interaction (\(0\)), and followship (\(+1\)) between two instances. This representation retains only the sign of each interaction, discarding the weight of positive ties, and allows us to treat the positive layer as time-invariant: a positive edge exists whenever at least one follower relationship is active, so the binary structure is robust to ordinary follower turnover and the follow network can be regarded as a stable backbone over the times studied.\\

\begin{figure*}
    \centering
    \includegraphics[width=0.995\textwidth]{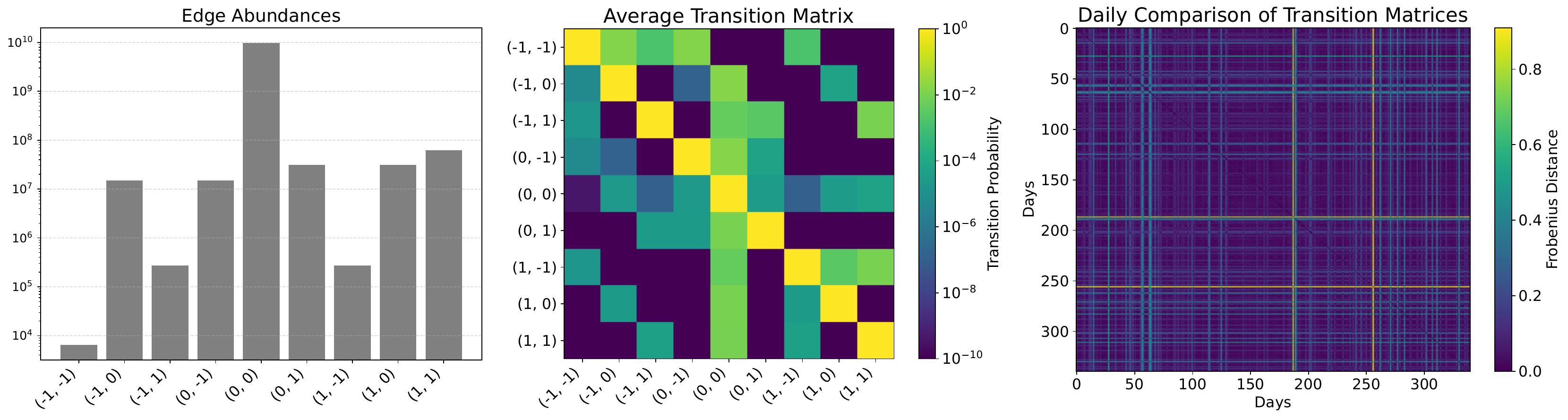}
    \caption{\textbf{Edge motif transitions and network stationarity.} Left: Distribution of dyadic edge configurations over the entire observation period. Center: Average daily transition matrix, showing strong diagonal dominance (state persistence). Right: Frobenius distance between daily transition matrices, confirming low temporal variability.}
    \label{fig:mastodon_edges}
\end{figure*}

\noindent
For each day \(t\), we construct a \(9 \times 9\) transition matrix recording how many dyads move from one configuration at time \(t\) to another at time \(t+1\). Row-normalizing these matrices yields daily probability transition matrices $P_t$, where entry $(P_t)_{ij}$ gives the probability that a dyad in state \(i\) transitions to state \(j\) in one step, with each row summing to one by construction. Figure~\ref{fig:mastodon_edges} (left panel) shows the distribution of dyadic configurations aggregated over the full observation period. As expected, the states \((0,0)\) and \((+1,+1)\) dominate, indicating that most instance pairs either remain unconnected or maintain reciprocal follow relations throughout the year. The average transition matrix (center panel) exhibits strong diagonal dominance: approximately \(98.6\%\) of dyads retain their configuration from one day to the next, confirming high local structural persistence at the dyadic level.\\
To quantify temporal variability in transition dynamics, we compute the Frobenius norm of pairwise differences between daily transition matrices. As shown in Figure~\ref{fig:mastodon_edges} (right panel), these distances remain consistently small throughout the year, with a maximum of approximately \(0.82\)—well below the theoretical upper bound of \(\sqrt{18} \approx 4.24\) for maximally divergent \(9 \times 9\) row-stochastic matrices. Similar conclusions hold using cosine distance between flattened matrices. Together, these results establish that edge reconfiguration events are both rare and highly consistent in character over time, despite the continuous turnover of instances participating in moderation.\\

\noindent
The strong persistence observed at the dyadic level suggests that the system may operate near a dynamical equilibrium. To test this, we model dyadic motif transitions as a discrete-time Markov process. Let $\bar{P}$ denote the (average) transition probability matrix and $\pi$ its stationary distribution. A stationary distribution must satisfy
\begin{equation}
\pi = \bar{P}^\top \pi,
\end{equation}
meaning the observed distribution remains invariant under the transition dynamics encoded in $\bar{P}$.\\

\begin{figure*}
    \centering
    \includegraphics[width=0.8\textwidth]{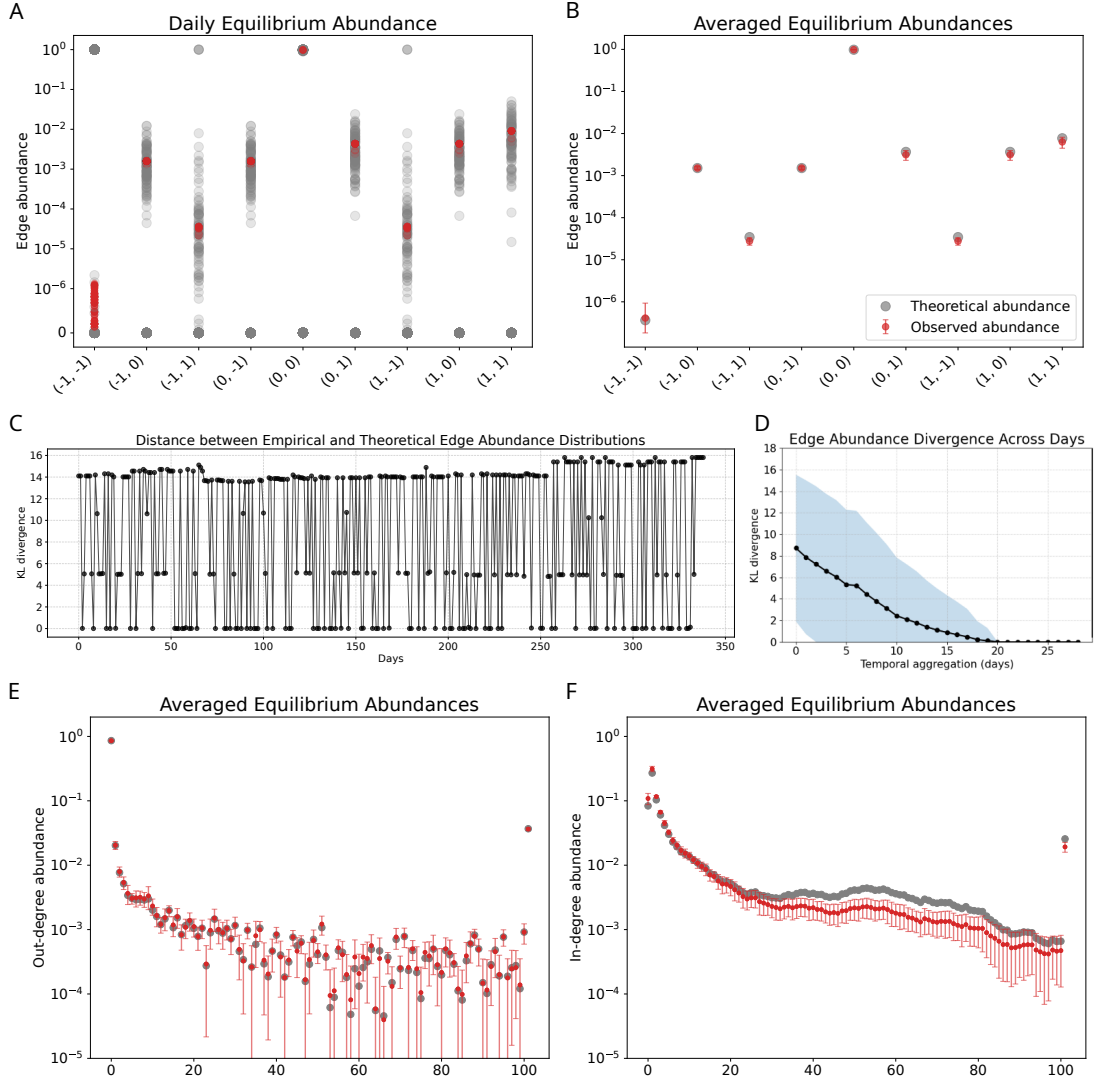}
    \caption{\textbf{Equilibrium analysis of edge and degree distributions.} 
    \textbf{A}: Daily empirical and theoretical edge motif abundances based on the stationarity condition. 
    \textbf{B}: Average edge motif abundances compared to the stationary distribution of the average transition matrix. 
    \textbf{C}: KL divergence between daily observed and theoretical edge abundances. 
    \textbf{D}: KL divergence decreases with temporal aggregation, reaching near-zero around 20 days. 
    \textbf{E}–\textbf{F}: Comparison between observed and predicted (equilibrium) out- and in-degree distributions based on the average transition matrix.}
    \label{fig:mastodon_equilibrium}
\end{figure*}

\noindent
We first evaluate this condition at daily resolution. For each day \(t\), we compute the predicted abundance vector $\bar{P}^\top \hat{\pi}_t$ and compare it to the empirical distribution $\hat{\pi}_{t+1}$. While dominant configurations such as \((0,0)\) and \((+1,+1)\) are well reproduced, a result partly expected given that positive ties are held fixed by construction, motifs involving negative ties exhibit systematic deviations. These discrepancies are quantified by the Kullback--Leibler (KL) divergence between predicted and observed distributions (Fig.~\ref{fig:mastodon_equilibrium}C), which fluctuates substantially at daily resolution. Strict stationarity therefore does not hold at this timescale.\\

\noindent
However, when transition matrices are aggregated over longer temporal windows, a markedly different picture emerges. Using the transition matrix averaged over the full year yields a stationary distribution that closely matches empirical motif abundances (Fig.~\ref{fig:mastodon_equilibrium}B). More systematically, we compute KL divergences between predicted and observed distributions for rolling aggregation windows ranging from 1 to 30 days. As shown in Fig.~\ref{fig:mastodon_equilibrium}D, divergence decreases monotonically with window size, approaching zero for windows exceeding approximately 20 days. Stationarity thus emerges as an effective property at mesoscopic temporal scales, consistent with equilibrium-like behavior reported in other evolving relational systems~\cite{MIGUEL2025EQUILIBRIUM}.\\

\noindent
We extend this analysis to node-level statistics by examining degree transitions in the moderation layer. For each instance, we track how its in-degree and out-degree in the negative network evolve from day \(t\) to day \(t+1\), grouping degrees above 100 into a single category to reduce fluctuations due to the long tail of the distributions. Aggregating daily degree transition matrices over time and computing the implied stationary distributions, we find close agreement with the empirical degree distributions for both in- and out-degree (Figs.~\ref{fig:mastodon_equilibrium}E--F). This indicates that, although individual instance degrees fluctuate day to day, their macroscopic distribution is consistent with an underlying stationary process when viewed at aggregated temporal scales. Importantly, this analysis relies exclusively on the moderation layer, which is fully dynamic, and is therefore not subject to the stationarity assumption imposed on the positive network.\\

\noindent
As a stronger test of equilibrium, we examine detailed balance at the level of dyadic motifs. For a reversible Markov process, detailed balance requires
\begin{equation}
\pi_i ({\bar{P}})_{ij} = \pi_j (\bar{P})_{ji},
\end{equation}
meaning that probability flux between any two states is symmetric. At daily resolution, this condition is not satisfied, reflecting asymmetries driven by the continuously changing identities of moderating and targeted instances. However, deviations decrease steadily under temporal aggregation: when transition matrices are averaged over windows of approximately two weeks or longer, detailed balance is closely approximated (Supplementary Fig.~3). This result reinforces the conclusion that, despite continuous participant turnover, the system approaches a reversible equilibrium regime when dynamics are coarse-grained over intermediate timescales.

\section{Modeling information transmission}
The decentralized architecture of Mastodon distributes moderation authority across independently governed instances, which can restrict federation by issuing domain blocks and thereby alter the effective pathways through which information circulates. The equilibrium analysis presented above reveals that this moderation activity produces a stable macroscopic network structure. A natural follow-up question is whether this structure has measurable consequences for information diffusion: does decentralized moderation merely regulate behavior, or does it also shape the efficiency and directionality of information transmission across the federation?\\

\noindent
Inspection of the moderation network reveals a marked asymmetry between instances that predominantly issue domain blocks and those that are primarily targeted. We exploit this asymmetry to partition $V^*$ into two groups. The minority group $m$ consists of instances that actively enact bans; the majority group $M$ comprises instances that are predominantly banned. Instances that both issue and receive bans are assigned to the minority group, reflecting their active governance role. This partition captures the functional distinction between moderating and moderated instances within the federated structure, and maps directly onto the structural asymmetry identified in the moderation layer.\\

\begin{figure}
    \centering
    \includegraphics[width=0.45\textwidth]{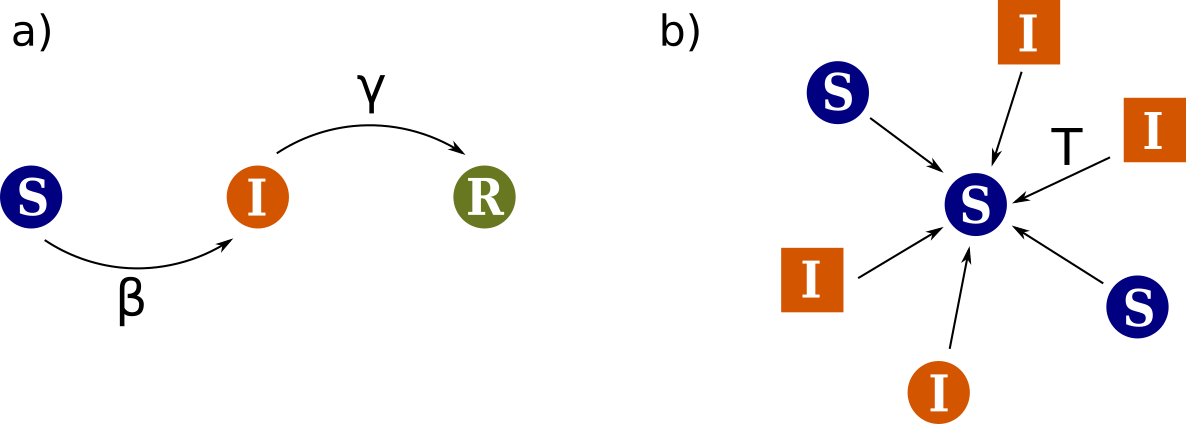}
    \caption{\textbf{Spreading model:} (a) Simple contagion: SIR compartmental model. (b) Complex Contagion, with threshold T. The central node will become informed if it has a fraction of informed neighbours, no matter which the belong, higher than T.}
    \label{fig:esquema_spreading}
\end{figure}

\noindent
To investigate how this structural partition shapes diffusion, we simulate information transmission on the aggregated positive (follow) network using a hybrid contagion framework that combines simple and complex transmission mechanisms (Fig.~\ref{fig:esquema_spreading}). The choice of a two-mechanism model is motivated by the sociological distinction between within-group and cross-group adoption: information spreading among structurally similar, well-connected instances can be expected to follow simple exposure dynamics, whereas adoption across group boundaries—where instances may differ substantially in norms and governance practices—typically requires greater social reinforcement \cite{CENTOLA2010OSNEXPERIMENT}.\\

\noindent
Within each group, information spreads via a discrete-time SIR model. A susceptible node \(i\) becomes informed with probability \(\beta\) upon contact with each informed neighbor, then transitions to the recovered state after one transmission attempt. Across group boundaries, we implement a complex contagion mechanism: a susceptible node adopts the information only if the fraction of its informed neighbors exceeds a threshold \(T \in [0,1]\), irrespective of which group those neighbors belong to. This threshold captures the additional reinforcement required for cross-group adoption.\\

\noindent
Simulations are performed on a static network obtained by averaging the daily moderation networks over a month, which is window consistent with the equilibrium timescale identified in the previous section, combined with the time-invariant positive layer. This choice ensures that the network used for diffusion simulations reflects the stable macroscopic configuration rather than transient daily fluctuations.\\

\noindent
For each simulation, we seed infection in a source group \(s \in \{M, m\}\) and measure the final fraction of informed nodes in a target group \(t \in \{M, m\}\), yielding four transmission scenarios:
\[
IT_{MM}, \quad IT_{Mm}, \quad IT_{mM}, \quad IT_{mm},
\]
where \(IT_{st}\) denotes the final attack rate in group \(t\) when the source lies in group \(s\).\\

\begin{figure}[h!]
    \centering
    \includegraphics[width=0.4\textwidth]{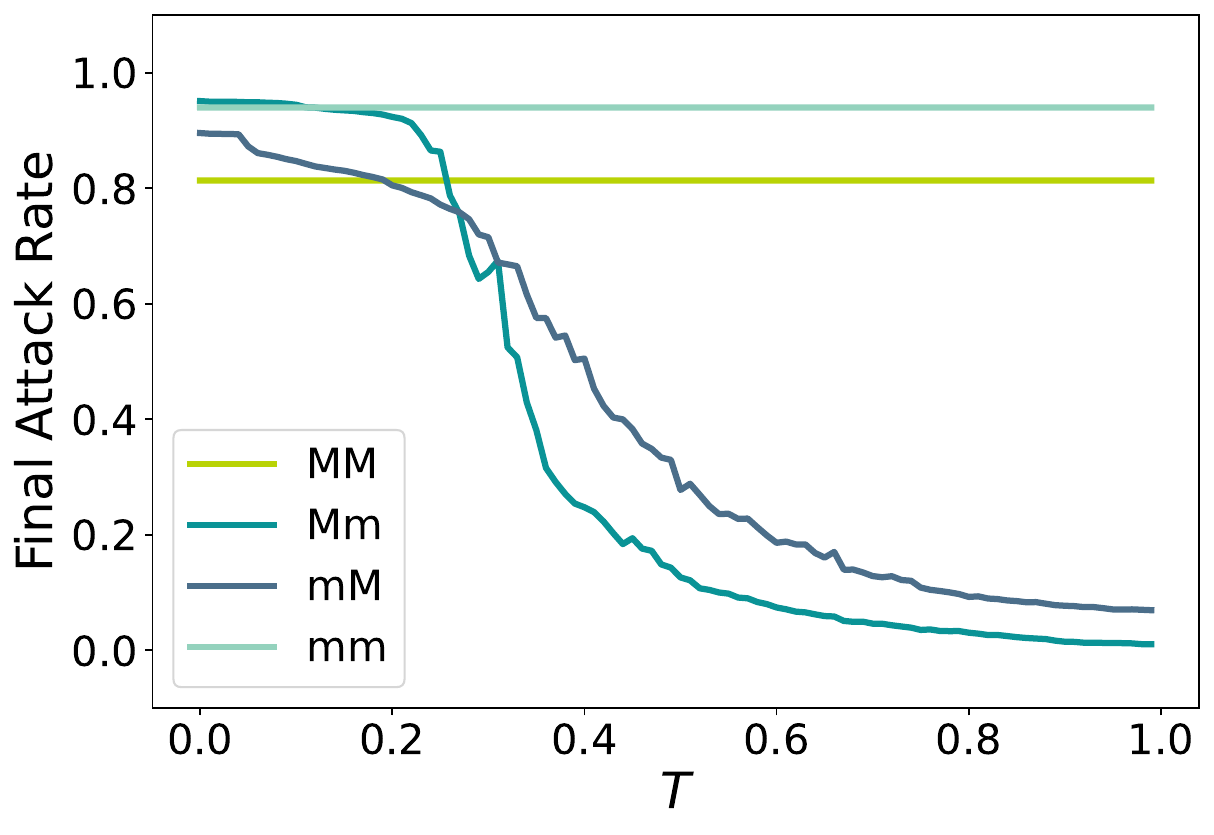}
    \caption{\textbf{Final attack rates for different source-target group combinations as a function of the cross-group adoption        threshold $T$.} Simulations are run on the governance-social interface with $\beta = 1.0$ and $\gamma = 1.0$.    }
    \label{fig:IT}
\end{figure}

\noindent
We first consider \(\beta = 1.0\) as an upper bound on within-group contagion. Figure~\ref{fig:IT} shows attack rates as a function of the threshold \(T\). When information originates in the minority group, within-group diffusion is nearly complete (\(IT_{mm} \approx 96\%\)), indicating strong internal cohesion. Within-group diffusion in the majority reaches approximately \(80\%\), suggesting comparatively weaker internal connectivity.\\
Cross-group transmission exhibits asymmetry: for \(T = 0\), diffusion is substantial in both directions, but as \(T\) increases, transmission from the majority to the minority (\(IT_{Mm}\)) declines faster, while transmission from the minority to the majority (\(IT_{mM}\)) remains more robust over a broader range of thresholds. The minority group is therefore structurally more effective at broadcasting information across group boundaries.\\

\begin{figure}[h!]
    \centering
    \includegraphics[width=0.5\textwidth]{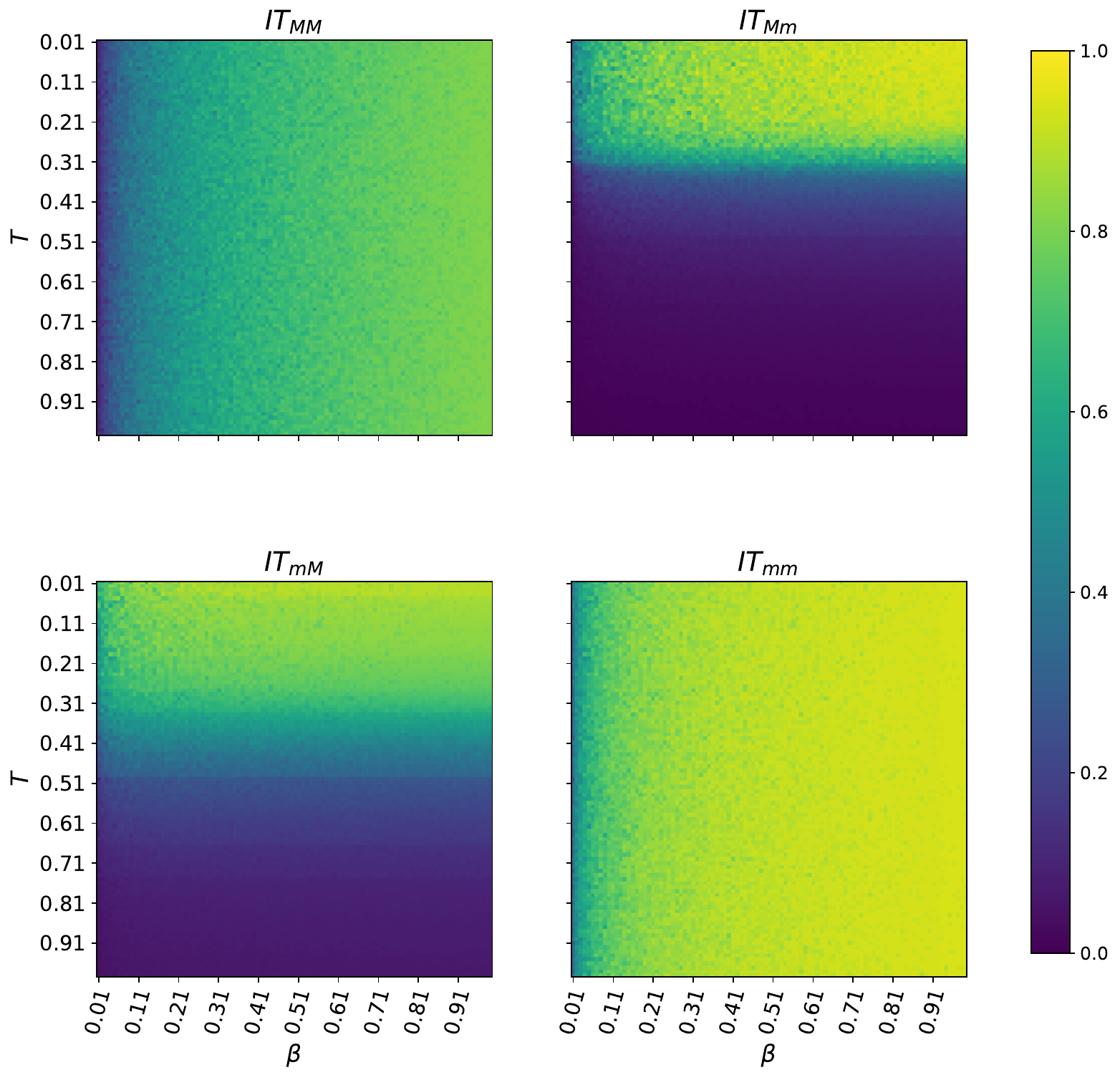}
    \caption{\textbf{Final attack rates for all source–target group combinations as a function of the threshold $T$, across varying values of infection probability $\beta$.} }
    \label{fig:IT_map}
\end{figure}

\noindent
Exploring the full joint parameter space of \(\beta\) and \(T\) (Fig.~\ref{fig:IT_map}) confirms these patterns across a wide range of contagion conditions. Within-minority diffusion remains efficient even at moderate \(\beta\), while cross-group transmission from the majority to the minority is consistently the most fragile pathway, highly sensitive to both threshold and contagion probability. These results suggest that the governance-active minority occupies a structurally advantageous position for information dissemination throughout the federation.\\

\noindent
To systematically characterize directional asymmetries across the full parameter space, we define four aggregate observables following~\cite{DIAZDIAZ2022BIAS}:
\begin{align}
\bar{IT} &= \frac{1}{4}\left(IT_{mm}+IT_{mM}+IT_{Mm}+IT_{MM}\right), \\
B_E &= \frac{1}{2}\left(-IT_{mm}-IT_{mM}+IT_{Mm}+IT_{MM}\right), \\
B_R &= \frac{1}{2}\left(-IT_{mm}+IT_{mM}-IT_{Mm}+IT_{MM}\right), \\
B_{EC} &= \frac{1}{2}\left(IT_{mm}-IT_{mM}-IT_{Mm}+IT_{MM}\right).
\end{align}

The mean transmission \(\bar{IT}\) measures overall diffusion efficiency across all scenarios. The emissivity bias \(B_E\) quantifies whether one group spreads information more effectively; a negative value indicates the minority is the more efficient emitter. The receptivity bias \(B_R\) captures asymmetries in receiving information. The echo chamber bias \(B_{EC}\) measures whether transmission occurs predominantly within groups (\(B_{EC} > 0\)) or across them (\(B_{EC} < 0\)).
\\

\begin{figure}[h!]
    \centering
    \includegraphics[width=0.5\textwidth]{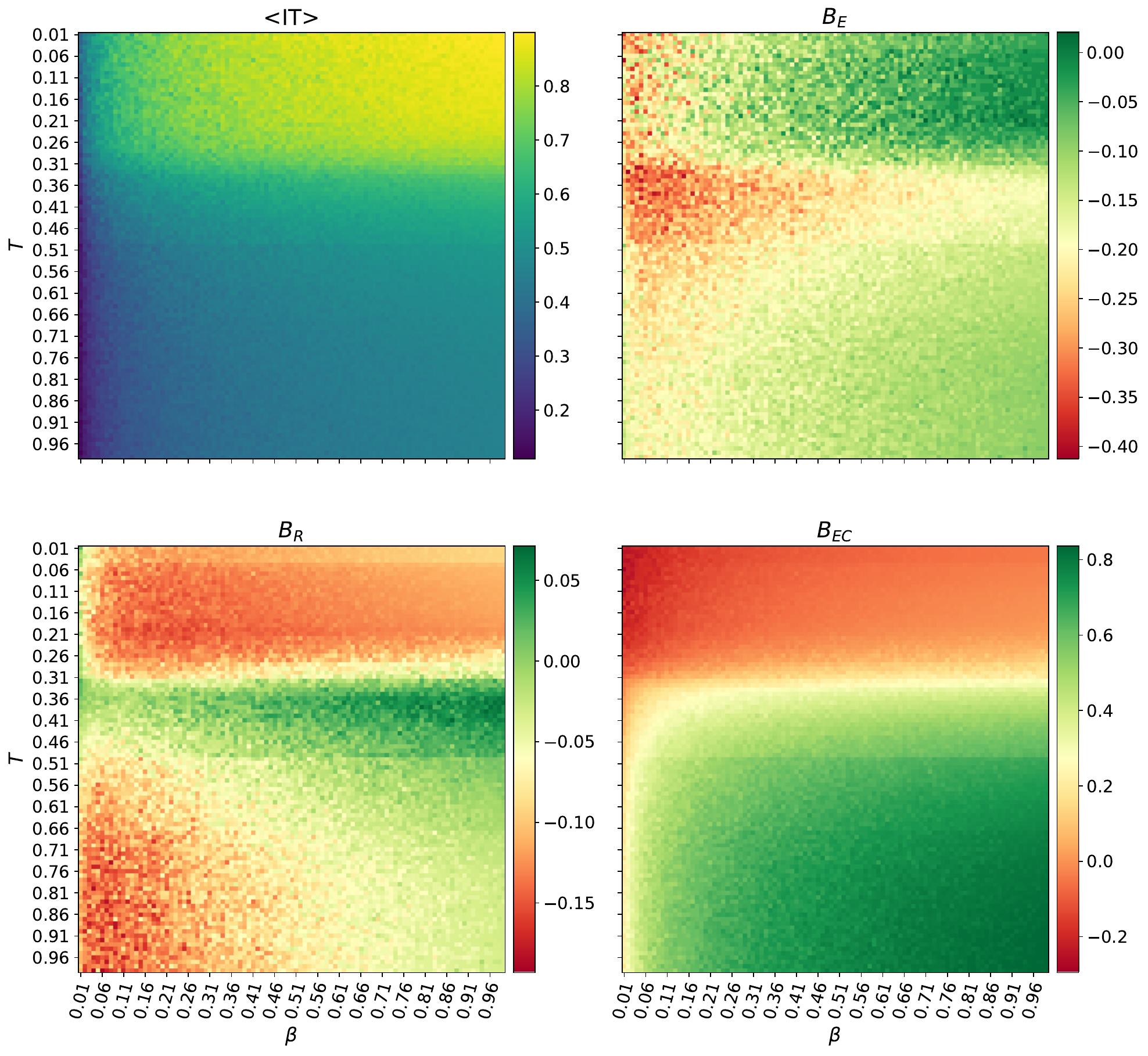}
    \caption{ \textbf{Information transmission biases as a function of the threshold $T$, across varying values of the contagion probability $\beta$.} The panels show four aggregated observables: $\langle IT \rangle$ — the average information transmission across all source–target combinations; $B_E$ — the emissivity bias, indicating whether one group is more efficient at spreading information; $B_R$ — the receptivity bias, indicating whether one group is more likely to receive information; $B_{EC}$ — the echo chamber bias, measuring the extent to which information remains confined within groups.}
    \label{fig:metrics}
\end{figure}

\noindent
Figure~\ref{fig:metrics} shows how these observables vary across \((\beta, T)\). As expected, \(\bar{IT}\) decreases with larger thresholds and lower contagion probabilities. Across most of the parameter space, \(B_E\) remains negative, confirming that the minority group is systematically the more efficient emitter of information. The receptivity bias \(B_R\) is comparatively small in magnitude and exhibits non-monotonic behavior, suggesting that receptivity asymmetries are secondary to emissivity effects. \\
The echo chamber bias \(B_{EC}\) reveals two distinct regimes: under permissive contagion conditions (low \(T\)), meaningful cross-group diffusion occurs, keeping \(B_{EC}\) moderate; under restrictive conditions (high \(T\)), \(B_{EC}\) approaches unity, indicating strong within-group confinement. Importantly, these echo-chamber effects emerge in a network that is globally structurally balanced, demonstrating that compartmentalization here is not a consequence of polarization but of the structural boundaries imposed by decentralized moderation.\\

\noindent
To quantify the contribution of moderation to these diffusion asymmetries, we simulate a loose moderation scenario in which domain block links—normally structural barriers to information flow—are reclassified as active positive connections, allowing information to propagate freely across previously banned pathways. Comparing this relaxed configuration to the baseline via differences in the aggregate observables isolates the effect of moderation structure on diffusion dynamics.\\

\begin{figure}[h!]
    \centering
    \includegraphics[width=0.4\textwidth]{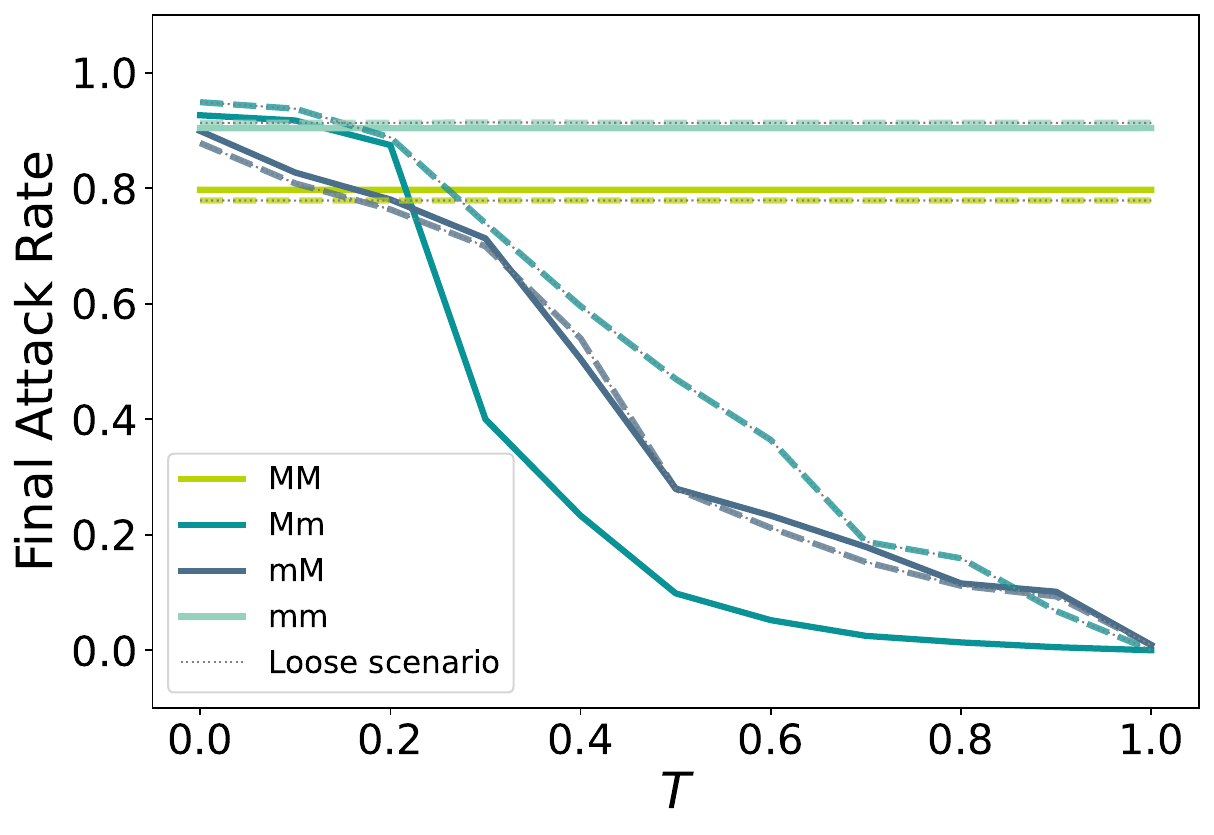}
    \caption{\textbf{Loose moderation scenario.} Final attack rates $IT_{st}$ for all source–target group combinations when banned links are treated as active communication channels. Compared with the baseline results in Figure~\ref{fig:IT}.}
    \label{fig:IT_loose}
\end{figure}

\begin{figure*}
    \centering
    \includegraphics[width=0.95\textwidth]{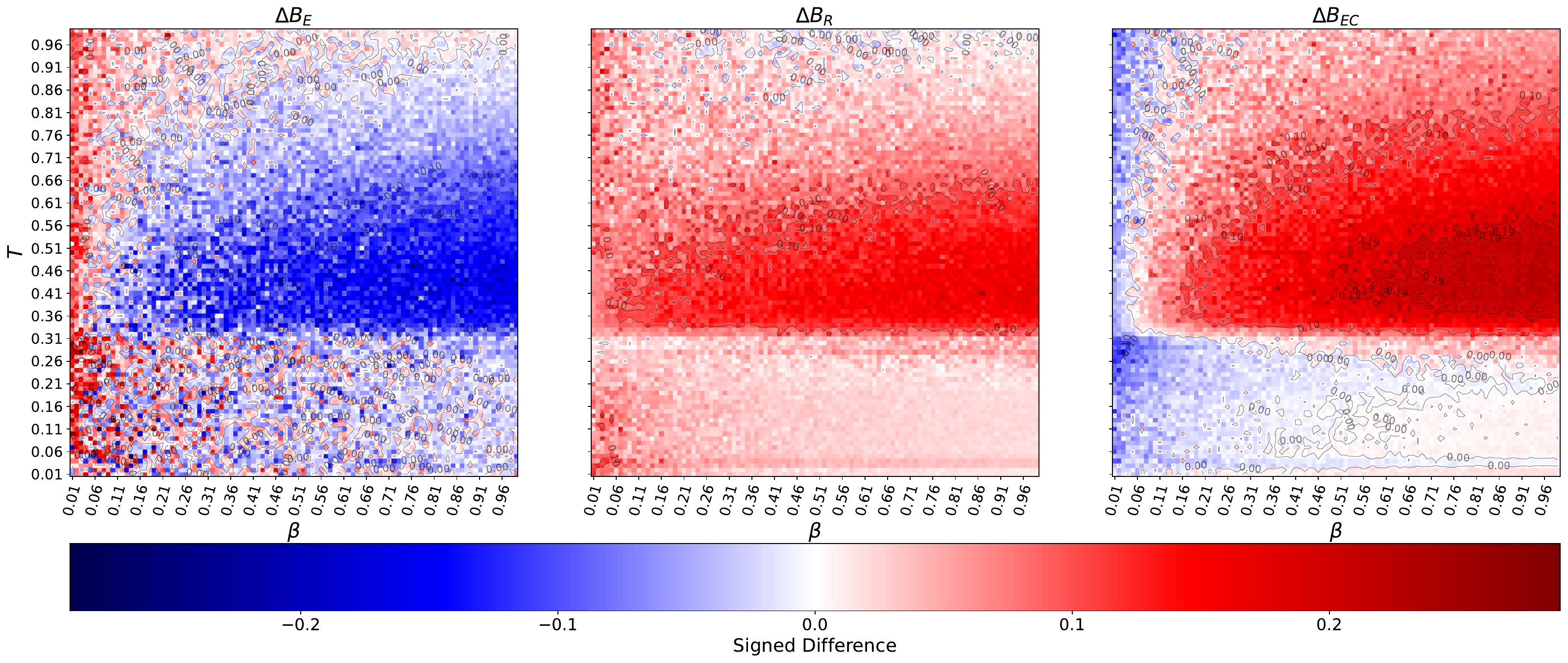}
    \caption{\textbf{Difference plots between baseline and loose moderation cases.} Each panel shows the signed difference of the respective observable ($\Delta \bar{IT}$, $\Delta B_E$, $\Delta B_R$, $\Delta B_{EC}$). Red regions indicate higher values in the baseline case; blue regions indicate higher values in the loose case.}
    \label{fig:metrics_diff}
\end{figure*}

\noindent
Figure~\ref{fig:IT_loose} shows that relaxing moderation increases overall spread, with cross-group transmission from the minority to the majority eventually surpassing the reverse direction (\(IT_{mM} > IT_{Mm}\)) as more pathways become available. The nuanced effects are captured in Figure~\ref{fig:metrics_diff}, which shows signed differences between baseline and loose observables. \\

\noindent
The emissivity bias \(B_E\) becomes more negative under relaxed moderation, indicating that lifting bans amplifies the minority's already superior broadcasting capacity—moderation therefore partially suppresses the outreach of the governing minority. The receptivity bias \(B_R\) is higher in the baseline, suggesting that structured moderation focuses transmission and enhances targeted reception in the majority, whereas relaxing it introduces diffusion noise. Finally, \(B_{EC}\) is stronger under baseline conditions, confirming that decentralized moderation actively reinforces within-group confinement. Even in the loose scenario, however, anti-echo-chamber effects remain weak and localized, indicating that cross-group transmission is structurally constrained beyond what moderation alone imposes—a consequence of the underlying topology of the follow network itself.\\

\section{Discussion}
The rise of decentralized online social networks (DOSNs) like Mastodon has introduced new challenges for understanding the dynamics of information spread and moderation in the absence of centralized control. While a substantial body of research has examined content diffusion and governance in traditional, centralized platforms, much less is known about how these processes unfold in federated systems composed of autonomous, self-governed instances. In this work, we investigated how decentralized moderation shapes both the structural stability and the information dynamics of Mastodon's inter-instance network over one year of observation. By combining a snapshot of inter-instance follow relationships with a longitudinal record of daily domain block actions, we constructed a signed, directed, temporal network at the instance level and analyzed its evolution through the lens of dynamical equilibrium and information diffusion.\\

\noindent
 Our first main finding is that despite continuous moderation activity and persistent turnover in the identities of moderating and moderated instances, the network exhibits robust macroscopic stability. Dyadic edge configurations transition rarely and consistently, degree distributions in the moderation layer are well described by stationary processes, and detailed balance is approximately satisfied when dynamics are coarse-grained over windows of approximately two weeks. This equilibrium-like behavior is remarkable given that no centralized authority coordinates moderation decisions: stability emerges from the aggregate effect of thousands of independent governance choices. This finding resonates with recent evidence of equilibrium-like properties in offline social networks~\cite{MIGUEL2025EQUILIBRIUM}, suggesting that the tendency toward stable macroscopic configurations may be a general feature of relational systems subject to sustained but locally governed perturbations, extending beyond personal relationships to digitally mediated environments.\\

\noindent
Our second main finding concerns the consequences of this stable structure for information transmission. The partition of instances into a governance-active minority and a moderated majority maps onto pronounced asymmetries in diffusion efficiency. Information originating in the minority spreads more effectively both within that group and outward toward the majority, while the reverse pathway is fragile and highly sensitive to contagion parameters. These asymmetries intensify under complex contagion conditions, where cross-group adoption requires reinforcement from multiple informed neighbors. The minority's structural advantage thus reflects not merely its smaller size or higher internal cohesion, but its privileged position within the follow network topology: moderating instances are better connected across the federation in ways that facilitate outward broadcast.\\
Instances that are widely banned are effectively isolated from the information flows originating in the governing minority, not because users self-select into ideological bubbles, but because domain blocks sever the pathways through which information would otherwise travel. The loose moderation scenario makes this mechanism explicit: relaxing bans increases overall diffusion and softens group boundaries, yet cross-group transmission remains constrained even without moderation barriers, pointing to the follow network topology as an independent contributor to compartmentalization.\\

\noindent
These findings suggest that decentralized moderation, far from being ungovernable, may represent a structurally self-organizing alternative to centralized content governance — one whose effectiveness emerges from collective action rather than institutional design. Without any centralized authority, the collective effect of independent instance-level block decisions produces a stable macroscopic configuration that isolates norm-violating domains from the broader information ecosystem. This stands in contrast to the mixed evidence on centralized deplatforming, which can reduce immediate harm but risks displacing harmful communities to less regulated spaces~\cite{ALI2021DEPLATFORMING, MEKACHER2023DEPLATFORMING}. In Mastodon, the federated structure means that banned instances are not removed from the platform entirely but are progressively excluded from the communication channels of governing instances — a softer but potentially more sustainable form of containment.\\

\noindent
Several limitations of this study point to productive directions for future work. First, our analysis operates exclusively at the instance level: we aggregate follower relationships and moderation actions across all users within each instance, which may obscure important individual-level heterogeneity. Second, the follow network is treated as a time-invariant backbone, an assumption we justify by the different timescales of the moderation and following processes, but which does not capture the gradual evolution of social ties over longer timescales. Incorporating temporal follow data would allow a more complete picture of how the positive and negative layers co-evolve. Third, our contagion model, while capturing the simple/complex distinction between within- and cross-group transmission, does not account for the semantic content of information or the heterogeneous receptivity of individual instances. Extending the framework to incorporate content-aware diffusion models could yield more refined predictions about which types of information are most affected by moderation boundaries.\\

\begin{acknowledgments}
B.A.G. thanks Dr. Fernando Díaz-Díaz and Miguel Ángel González-Casado for their valuable discussions and insightful feedback.
B.A.G, A.B., A.T. and R.G. acknowledge the financial support received from the European Union’s Horizon Europe research and innovation programme under grant agreement no. 101135437.
S.M. acknowledges financial support by the Spanish State Research Agency
(MICIU/AEI/10.13039/501100011033) and FEDER (UE) under project COSASTI
(PID2024-157493NB-C22), and the María de Maeztu project
CEX2021-001164-M.
\end{acknowledgments}

\bibliography{refs}

\end{document}